\documentstyle[aps]{revtex}
\begin{document}
\bibliographystyle{prsty}
\baselineskip=8.5mm
\parindent=7mm
\begin{center}
{\Large {\bf \sc{ Structures of the $f_0(980)$, $a_0(980)$ mesons and the strong coupling
constants $g_{f_0 K^+ K^-}$ , $g_{a_0 K^+ K^-}$ with
  the light-cone QCD sum rules }}} \\[2mm]
Zhi-Gang Wang$^{1,2}$ \footnote{wangzgyiti@yahoo.com.cn .},
Wei-Min Yang$^{2,3}$ and Shao-Long Wan$^{2,3} $    \\
$^{1}$ Department of Physics, North China Electric Power University, Baoding 071003, P. R. China \footnote{Mailing address.}\\
$^{2}$ CCAST (World Laboratory), P.O. Box 8730, Beijing 100080,
P. R. China \\
$^{3}$ Department of Modern Physics, University of Science and Technology of China, Hefei 230026, P. R. China \\
\end{center}

\begin{abstract}
In this article, with the assumption of explicit isospin   violation arising from the $f_0(980)-a_0(980)$ mixing,
we take the point of view that the scalar mesons $f_0(980)$ and
$a_0(980)$ have both strange and non-strange quark-antiquark
components  and evaluate the strong coupling constants $g_{f_0 K^+
K^-}$ and $g_{a_0 K^+ K^-}$  within the framework of the light-cone QCD sum rules
approach. The large strong scalar-$KK$
couplings through both the $n\bar{n}$ and $s\bar{s}$ components
$g^{\bar{n}n}_{f_0 K^+ K^-}$, $g^{\bar{s}s}_{f_0 K^+ K^-}$,
 $g^{\bar{n}n}_{a_0 K^+ K^-}$ and $g^{\bar{s}s}_{a_0 K^+ K^-}$will support the
hadronic dressing mechanism, furthermore, in spite of the constituent
structure differences  between the $f_0(980)$ and $a_0(980)$
mesons, the strange components have larger strong coupling
constants with the $K^+K^-$ state than the corresponding non-strange
ones, $g_{f_0 K^+ K^-}^{\bar{s}s}\approx \sqrt{2}g_{f_0 K^+
K^-}^{\bar{n}n}$ and $g_{a_0 K^+ K^-}^{\bar{s}s}\approx \sqrt{2}
g_{a_0 K^+ K^-}^{\bar{n}n}$. From the existing controversial values,
we can not reach  a general consensus on the strong coupling constants  $g_{f_0 K^+
K^-}$, $g_{a_0 K^+ K^-}$ and the mixing angles.
\end{abstract}

PACS numbers:  12.38.Lg; 13.25.Jx; 14.40.Cs

{\bf{Key Words:}}  $f_0(980)$, $a_0(980)$ , light-cone QCD sum
rules
\section{Introduction}
The constituent quark model provides a rather successful
description of the spectrum of the mesons in terms of
quark-antiquark bound states, which fit into the suitable
multiplets reasonably well. However, the scalar mesons present a
remarkable exception as the structures of those mesons are not
unambiguously determined yet \cite{Godfray,Close2002}. From the
point of view of experiment, the broad width
( for the $f_0(980)$, $a_0(980)$ et al, the widths are comparatively narrow) and strong overlaps
with the continuum background make those particles difficult to
resolve. On the other hand, the numerous candidates  with the same
quantum numbers for the  quark-antiquark $(q{\bar q})$  scalar
states obviously exceed the prediction power of the constituent
quark model in the energy region below 2 GeV, for example, the
isospin $I=0$ scalars $f_0(400-1200)$,  $f_0(980)$, $f_0(1370)$,
$f_0(1500)$ and $f_0(1710)$ can not be accommodated in one
$q\bar{q}$ nonet,  some are supposed to be glueball, molecule,
multi-quark state, etc. In fact, the light scalar mesons are the
subject of an intense and continuous controversy in clarifying the
hadron spectroscopy, the more elusive things are the constituent
structures  of the $f_0(980)$ and $a_0(980)$ mesons with almost
the degenerate masses. In the naive constituent quark model, the
isovector $a_0(980)$ meson is interpreted as
$a_0=(u\overline{u}-d\overline{d})/\sqrt{2}$ and the isoscalar
$f_0(980)$ meson is taken as pure $s\bar{s}$ state $f_0=s\overline{s}$; while the four quark
$qq{\overline{qq}}$ state suggestions propose that the $f_0(980)$
and $a_0(980)$ mesons could either
 be compact  objects i.e. nucleon-like bound states of
quarks  with symbolic quark structures  $f_0={s{\overline s}({ u
{\overline u}+d {\overline d})/ \sqrt{2}}}$ and $a_0=s {\overline
s}( u {\overline u}-d {\overline d}) / \sqrt{2}$
\cite{JaffeAchasov4q}, or spatially  extended objects i.e. deuteron-like
bound states of hadrons, for example, the $f_0(980)$ meson is usually
taken as a $K {\overline K}$ molecule, etc \cite{IsgurKK}. The
hadronic dressing mechanism takes the point of view that the $f_0(980)$ and
$a_0(980)$ mesons have small $q\bar{q}$ cores of typical $q\bar{q}$
meson size, the strong couplings to the hadronic channels  enrich the pure $q\bar{q}$
states with other components and spend part (or most part) of
their lifetime as virtual $ K \bar{K} $ states \cite{HDress}. Despite
 what constituents they may have, we have the fact that  they both lie just a little
below the $K\bar{K}$ threshold, the  strong interactions with
the $K\bar{K}$ threshold will significantly influence their dynamics.
In strong interactions (QCD), the isospin is believed to be a nearly exact
symmetry, broken only by the slight masses  difference between the
$u$ and $d$ quarks, or electroweak effects, however, the
 mass gaps between the $f_0(980)$,
$a_0(980)$ and the $K^+K^-$
and $K^0\bar{K^0}$ thresholds make an exception and can not
be explained. The mixing of the
two scalar mesons i.e. the isospin broken can occur through the transitions between the
intermediate $K^+K^-$, $K^0\bar{K^0}$ states. In
Ref.\cite{CloseKirk00}, analysis of the central production
in the reaction $pp\rightarrow p_s(\eta\pi^0)p_f$
shows  that the $f_0(980)$ and
$a_0(980)$ mesons can mix substantially with each other with
intensity about $(8\pm3)\%$ and the isospin symmetry is obviously
broken. The isospin mixing effects could considerably alter some existing predictions for the radiative
decays $\phi\rightarrow f_0\gamma$ and $\phi\rightarrow a_0\gamma$\cite{CloseKirk01},  however,
further studies show that when the physical masses and widths are included, the
mixing effects are very small \cite{Achasov02}; the  vector meson dominance model
also indicates the mixing effects are small \cite{VMD02}.
On the other hand, the generalized J\"{u}lich meson exchange model for $\pi\pi,K\bar{K},\pi\eta $
 scattering with
physical mass eigenstates predicts that the charged and neutral K mass splitting induced isospin violation
and the coupled $\pi\pi -K\bar{K} $
channels induced G-parity violation give rise to nonvanishing cross section for the $\pi\pi-\pi^0 \eta $ transition and lead to
  the $f_0(980)-a_0(980)$ mixing \cite{KRS97}. In Ref.\cite{NNAchasov04}, the authors suggest that
perform the polarized target experiments on the reaction $\pi^-p\to\eta\pi^0n$ at high energy
in which the existence of $a^0_0(980)-f_0(980)$ mixing can be unambiguously and very easily established
through the presence of a strong jump in the azimuthal asymmetry of the $\eta\pi^0$ $S$ wave production
cross section near the $K\bar K$ thresholds.
If we take the $f_0(980)-a_0(980)$ mixing and explicit isospin violation  for granted, no matter how tiny they are,
we can  take the point of
view that the $f_0(980)$ and $a_0(980)$ mesons both have two possible constituent $q\bar{q}$ states i.e.
$n {\bar n}$ and $s {\bar s}$ in the $q\bar{q}$ quark model; in the isospin limit, the $a_0(980)$ meson
has pure $n\bar{n}$ quark structure $a_0=\frac{u\bar{u}-d\bar{d}}{\sqrt{2}}$ with
isospin $I=1$ and can not have $s\bar{s}$ component while the isospin $I=0$ meson $f_0(980)$
can have both $n {\bar n}=\frac{u\bar{u}+d\bar{d}}{\sqrt{2}}$ and $s {\bar s}$ components. The $K$-matrix analysis of
the channels $f_0\rightarrow \pi\pi,\pi\pi\pi\pi,K\bar{K},\eta\eta,\eta\eta'$ shows that
the $f_0(980)$ meson may have both $n {\bar n}$ and $s {\bar s}$ components, even gluonium component\cite{Anisovich}.
 The radiative decays of the $\phi(1020)$ meson
$\phi(\rightarrow K^+K^-)\rightarrow a_0\gamma\rightarrow\gamma\pi\eta$ and $\phi(\rightarrow K^+K^-)\rightarrow f_0\gamma\rightarrow\gamma\pi\pi$
provide an efficient tool to investigate  the structures of  the
$a_0(980)$ and $f_0(980)$   mesons. It is generally agreed
that the experimental data supports the $K\bar{K}$  mesons  loop
mechanism for those decays,  where the radiative decays occur
through the photon emission from the intermediate $K^+K^-$ loop.
The important hadronic parameters entering the analysis involving
the $f_0(980)$   and $a_0(980)$ mesons  are   the strong coupling
constants $g_{f_0 K^+ K^-}   $ and $ g_{a_0 K^+ K^-}  $ .

In this article, we take the point of view that the $f_0(980)$ and $a_0(980)$ mesons
are mixed states which consist of both $n {\bar n}$ and $s {\bar s}$ components, and devote to
determine the values of the strong coupling constants  $g_{f_0 K^+
K^-}$ and $g_{a_0 K^+ K^-}$ within the framework of  the light-cone QCD sum rules approach,
which carries out the
 operator product expansion  near the light-cone $x^2\approx 0$
 instead of the short distance $x\approx 0$ while the nonperturbative
 matrix elements  are parameterized by the light-cone distribution amplitudes
 which classified according to their twists  instead of
 the vacuum condensates \cite{Balitsky,Braun1,Chernyak1}.

The article is arranged as: in Section II, the strong coupling
constants  $g_{f_0 K^+ K^-}$ and $g_{a_0 K^+ K^-}$ are evaluated
with the light-cone QCD sum rules approach; and in Section III,
conclusion.

\section{strong Coupling constants  $g_{f_0 K^+ K^-}$ and $g_{a_0 K^+ K^-}$ with light-cone QCD sum rules}

In the following, we write down the definitions  for the strong
coupling constants $g_{f_0 K^+ K^-}$ and $g_{a_0 K^+ K^-}$,
\begin{eqnarray}
\langle K^+(q) K^-(p)|f_0(p+q) \rangle=g_{f_0 K^+ K^-}, \ \ \
\langle K^+(q) K^-(p)|a_0(p+q) \rangle=g_{a_0 K^+ K^-}.
\end{eqnarray}
In this article, we investigate the strong coupling constants
$g_{f_0 K^+ K^-}$ and $g_{a_0 K^+ K^-}$ with  the scalar
interpolating currents $J_{f_0}$ and $J_{a_0}$,
and choose the following two two-point correlation functions,
\begin{eqnarray}
&& J_{f_0}=sin\theta \frac{\bar{u}u+\bar{d}d}{\sqrt{2}}+cos\theta \bar{s}s, \\
&& J_{a_0}=sin\varphi \frac{\bar{u}u-\bar{d}d}{\sqrt{2}}+cos\varphi \bar{s}s,
\end{eqnarray}
\begin{eqnarray}
&&T^{f_0}_{\mu}(p,q)=i \int d^4x \, e^{i p \cdot x} \,
\langle{K^+(q)}|T[J_\mu(x) J_{f_0}(0)]|0\rangle , \\
&&T^{a_0}_{\mu}(p,q)=i \int d^4x \, e^{i p \cdot x} \,
\langle{K^+(q)}|T[J_\mu(x) J_{a_0}(0)]|0\rangle.
\end{eqnarray}
Here the axial-vector
current  $J_\mu={\bar u}\gamma_\mu \gamma_5 s$ interpolates the
pseudoscalar K meson, and the external K state has four
momentum $q$ with $q^2=M_K^2$.
If the isospin violation  is small, the parameter $\varphi$ is close to $\frac{\pi}{2}$.
Those correlation functions in Eqs.(4-5) can be decomposed as
\begin{eqnarray}
&&T^{f_0}_{\mu}(p,q)=T^{f_0}_{p}\left(p^2,(p+q)^2\right)p_{\mu}+T^{f_0}_{q}
\left(p^2,(p+q)^2\right)q_{\mu},   \\
&&T^{a_0}_{\mu}(p,q)=T^{a_0}_{p}\left(p^2,(p+q)^2\right)p_{\mu}+T^{a_0}_{q}
\left(p^2,(p+q)^2\right)q_{\mu}
\end{eqnarray}
due to the tensor analysis.

With the basic assumption of hadron-quark duality of the QCD sum
rules approach \cite{Shifman79}, we can insert  a complete series of
intermediate states with the same quantum numbers as the current
operators $J_{f_0}$ , $J_{a_0}$ and $J_\mu$ into those
correlation functions in Eqs.(4-5) to obtain the hadronic representation. After
isolating the ground state contributions from the pole terms of
the  $f_0(980)$, $a_0(980)$ and K mesons, we get
the following result,
\begin{eqnarray}
T^{f_0}_p\left(p^2,(p+q)^2\right)p_\mu&=&\frac{<0\mid
J_{\mu}\mid K(p)><KK\mid f_0> <f_0(p+q)|J_{f_0}\mid 0>}
  {\left(M_K^2-p^2\right)\left[M_{f_0}^2-(p+q)^2\right]} + \cdots \nonumber \\
&=&\frac{if_{K} g_{f_0K^+K^-} f_{f_0}M_{f_0}  p_\mu}
  {\left(M_K^2-p^2\right)\left[M_{f_0}^2-(p+q)^2\right]} + \cdots ; \\
T^{a_0}_p\left(p^2,(p+q)^2\right)p_\mu&=&\frac{<0\mid
J_{\mu}\mid K(p)><KK\mid a_0> <a_0(p+q)|J_{a_0}\mid 0>}
  {\left(M_K^2-p^2\right)\left[M_{a_0}^2-(p+q)^2\right]} + \cdots \nonumber \\
&=&\frac{if_{K} g_{a_0K^+K^-} f_{a_0}M_{a_0} p_\mu}
  {\left(M_K^2-p^2\right)\left[M_{a_0}^2-(p+q)^2\right]} + \cdots,
\end{eqnarray}
where the following definitions have been used,
\begin{eqnarray}
<f_0(p+q)\mid J_{f_0}\mid 0>&=&f_{f_0}M_{f_0}\,, \nonumber\\
<a_0(p+q)\mid J_{a_0}\mid 0>&=&f_{a_0}M_{a_0}\,, \nonumber\\
<0\mid J_{\mu}\mid K(p)>&=&if_Kp_\mu~~.
\end{eqnarray}
Here we have not shown the contributions from the higher
resonances and  continuum states explicitly as they are suppressed due to the
Borel transformation. In the ground state approximation, the
tensor structures  $T^{f_0}_q\left(p^2,(p+q)^2\right) q_{\mu}$ and
$T^{a_0}_q\left(p^2,(p+q)^2\right) q_{\mu}$ have no
contributions and neglected.

In the following, we briefly outline the  operator product expansion for the
correlation functions in Eqs.(4-5) in perturbative QCD theory.
The calculations are performed at the
large space-like momentum regions $(p+q)^2\ll 0$  and  $p^2\ll 0$,
which correspond to the small light-cone distance $x^2\approx 0$
required by the validity of the operator product expansion approach.
Firstly, let us  write down the propagator of a massive quark in the external gluon field
in the Fock-Schwinger gauge\cite{Belyaev:1994zk,BK},
\begin{eqnarray}
\langle 0 | T \{q_i(x_1)\, \bar{q}_j(x_2)\}| 0 \rangle &=&
 i \int\frac{d^4k}{(2\pi)^4}e^{-ik(x_1-x_2)}\Bigg\{
\frac{\not\!k +m}{k^2-m^2} \delta_{ij}
-\int\limits_0^1 dv\,  g_s \, G^{\mu\nu}_a(vx_1+(1-v)x_2)
\left (\frac{\lambda^a}{2} \right )_{ij}
\nonumber
\\
& &  \Big[ \frac12 \frac {\not\!k +m}{(k^2-m^2)^2}\sigma_{\mu\nu} -
\frac1{k^2-m^2}v(x_1-x_2)_\mu \gamma_\nu \Big]\Bigg\}\, ,
\end{eqnarray}
here $G^{\mu \nu }_a$ is the gluonic field strength, $g_s$ denotes the strong
coupling constant.
Substituting the above $u$ , $s$ quark propagators and the
corresponding K meson light-cone distribution amplitudes into  Eqs.(4-5) and
completing the integrals over $x$ and $k$, finally we obtain
\begin{eqnarray}
T^{f_0}_p(p^2,(p+q)^2)&=&sin\theta\frac{1}{\sqrt{2}}\left\{if_K \int_0^1 du\left\{ {M_K^2 \over m_s}
\varphi_p (u) {1 \over -(p+uq)^2}
-2{M_K^2 \over 6 m_s} \varphi_\sigma(u)(p
\cdot
q+u M_K^2)  {1 \over [-(p+uq)^2]^2} \right\}\right. \nonumber \\
&&\left.+if_{3K} M_K^2 \int_0^1 dv \left(2v-3 \right) \int
{\cal D}\alpha_i \varphi_{3K}(\alpha_i) {1 \over \{
[p+q(\alpha_1+v\alpha_3)]^2 \}^2 } \right\} \nonumber \\
&&+cos\theta\left\{if_K \int_0^1 du\left\{ {M_K^2 \over m_s}
\varphi_p (u) {1 \over m_s^2-(p+uq)^2}\right.\right. \nonumber\\
&&\left.\left.-2\left[ m_s g_2(u)+{M_K^2 \over 6 m_s} \varphi_\sigma(u)(p
\cdot
q+u M_K^2) \right] {1 \over [m_s^2-(p+uq)^2]^2} \right\} \right.\nonumber \\
&&\left.+if_{3K} M_K^2 \int_0^1 dv \left(2v-3 \right) \int
{\cal D}\alpha_i \varphi_{3K}(\alpha_i) {1 \over \{
[p+q(\alpha_1+v\alpha_3)]^2-m_s^2 \}^2 }  \right. \nonumber \\
&&\left.-4 if_K m_s M_K^2 \left\{ \int_0^1 dv (v-1) \int_0^1 d \alpha_3 \int_0^{\alpha_3} d \beta \int_0^{1-\beta}d \alpha
{\Phi(\alpha,1-\alpha-\beta,\beta)\over \{[p+(1-\alpha_3+v\alpha_3)q]^2-m_s^2\}^3}\right.\right.
\nonumber \\
&& \left.\left.+ \int_0^1 dv  \int_0^1 d \alpha_3 \int_0^{1-\alpha_3} d \alpha_1\int_0^{\alpha_1}d \alpha
{\Phi(\alpha,1-\alpha-\alpha_3,\alpha_3) \over \{[p+(\alpha_1+v \alpha_3)q]^2-m_s^2\}^3
} \right\} \right\};
\end{eqnarray}
\begin{eqnarray}
T^{a_0}_p(p^2,(p+q)^2)&=&sin\varphi\frac{1}{\sqrt{2}}\left\{if_K \int_0^1 du\left\{ {M_K^2 \over m_s}
\varphi_p (u) {1 \over -(p+uq)^2}
-2{M_K^2 \over 6 m_s} \varphi_\sigma(u)(p
\cdot
q+u M_K^2)  {1 \over [-(p+uq)^2]^2} \right\}\right. \nonumber \\
&&\left.+if_{3K} M_K^2 \int_0^1 dv \left(2v-3 \right) \int
{\cal D}\alpha_i \varphi_{3K}(\alpha_i) {1 \over \{
[p+q(\alpha_1+v\alpha_3)]^2 \}^2 } \right\} \nonumber \\
&&+cos\varphi\left\{if_K \int_0^1 du\left\{ {M_K^2 \over m_s}
\varphi_p (u) {1 \over m_s^2-(p+uq)^2}\right. \right. \nonumber\\
&&\left.\left.-2\left[ m_s g_2(u)+{M_K^2 \over 6 m_s} \varphi_\sigma(u)(p
\cdot
q+u M_K^2) \right] {1 \over [m_s^2-(p+uq)^2]^2} \right\} \right.\nonumber \\
&&\left.+if_{3K} M_K^2 \int_0^1 dv \left(2v-3 \right) \int
{\cal D}\alpha_i \varphi_{3K}(\alpha_i) {1 \over \{
[p+q(\alpha_1+v\alpha_3)]^2-m_s^2 \}^2 }  \right. \nonumber \\
&&\left.-4 if_K m_s M_K^2 \left\{ \int_0^1 dv (v-1) \int_0^1 d \alpha_3 \int_0^{\alpha_3} d \beta \int_0^{1-\beta}d \alpha
{\Phi(\alpha,1-\alpha-\beta,\beta)\over \{[p+(1-\alpha_3+v\alpha_3)q]^2-m_s^2\}^3}\right.\right.
\nonumber \\
&& \left.\left.+ \int_0^1 dv  \int_0^1 d \alpha_3 \int_0^{1-\alpha_3} d \alpha_1\int_0^{\alpha_1}d \alpha
{\Phi(\alpha,1-\alpha-\alpha_3,\alpha_3) \over \{[p+(\alpha_1+v \alpha_3)q]^2-m_s^2\}^3
} \right\} \right\}.
\end{eqnarray}
In the limit $\theta=0$, our results for the expressions of  the three-particle twist-3
and twist-4 terms in Eq.(12) are slightly different from the  corresponding
ones in Ref.\cite{ColangeloFazio2003}, there may be some errors (or just writing errors) in
their calculations.
If we take the limit $\varphi=\frac{\pi}{2}$ in Eq.(13), the results for
 the isospin-vector scalar current
 are found, the expressions for the contributions from the
three-particle twist-3 light-cone distribution amplitudes may have
some errors (or just writing errors) in Ref.\cite{Yilmaz}.
However, the contributions from those terms are small and can not significantly
affect the numerical values.
Comparing with the mass of the $s$ quark, the masses of the $u$ and $d$ quarks
are neglected.

In calculation, the following  two-particle and three-particle
K meson light-cone distribution amplitudes are useful,
\begin{eqnarray}
<K(q)| {\bar u} (x) \gamma_\mu \gamma_5 s(0) |0>& =& -i f_K q_\mu
\int_0^1 du \; e^{i u q \cdot x} [\varphi_K(u)+x^2 g_1(u)] +f_K
\left(x_\mu - {q_\mu x^2 \over q \cdot x}\right)
\int_0^1 du \; e^{i u q \cdot x} g_2(u)  , \nonumber\\
<K(q)| {\bar u} (x) i \gamma_5 s(0) |0> &=& {f_K M_K^2 \over m_s}
\int_0^1 du \; e^{i u q \cdot x} \varphi_p(u)  \hskip 3 pt ,  \nonumber\\
<K(q)| {\bar u} (x) \sigma_{\mu \nu} \gamma_5 s(0) |0> &=&i(q_\mu
x_\nu-q_\nu x_\mu)  {f_K M_K^2 \over 6 m_s} \int_0^1 du \;
e^{i u q \cdot x} \varphi_\sigma(u),  \nonumber\\
<K(q)| {\bar u} (x) \sigma_{\alpha \beta} \gamma_5
g_s G_{\mu \nu}(v x)s(0) |0>&=&i f_{3 K}[(q_\mu q_\alpha g_{\nu \beta}-q_\nu q_\alpha g_{\mu
\beta}) -(q_\mu q_\beta g_{\nu \alpha}-q_\nu q_\beta g_{\mu
\alpha})] \int {\cal D}\alpha_i \; \varphi_{3 K} (\alpha_i)
e^{iq \cdot x(\alpha_1+v \alpha_3)} ,\nonumber\\
<K(q)| {\bar u} (x) \gamma_{\mu} \gamma_5 g_s G_{\alpha
\beta}(vx)s(0) |0>&=&f_K \Big[ q_{\beta} \Big( g_{\alpha \mu}-{x_{\alpha}q_{\mu}
\over q \cdot x} \Big) -q_{\alpha} \Big( g_{\beta
\mu}-{x_{\beta}q_{\mu} \over q \cdot x} \Big) \Big] \int {\cal{D}}
\alpha_i \varphi_{\bot}(\alpha_i)
e^{iq \cdot x(\alpha_1 +v \alpha_3)}\nonumber \\
&&+f_K {q_{\mu} \over q \cdot x } (q_{\alpha} x_{\beta}-q_{\beta}
x_{\alpha}) \int {\cal{D}} \alpha_i \varphi_{\|} (\alpha_i)
e^{iq \cdot x(\alpha_1 +v \alpha_3)} \hskip 3 pt ,  \nonumber\\
<K(q)| {\bar u} (x) \gamma_{\mu}  g_s \tilde G_{\alpha
\beta}(vx)s(0) |0>&=
&i f_K \Big[ q_{\beta} \Big( g_{\alpha \mu}-{x_{\alpha}q_{\mu}
\over q \cdot x} \Big) -q_{\alpha} \Big( g_{\beta
\mu}-{x_{\beta}q_{\mu} \over q \cdot x} \Big) \Big] \int {\cal{D}}
\alpha_i \tilde \varphi_{\bot}(\alpha_i)
e^{iq\cdot x(\alpha_1 +v \alpha_3)}\nonumber \\
&&+i f_K {q_{\mu} \over q \cdot x } (q_{\alpha}
x_{\beta}-q_{\beta} x_{\alpha}) \int {\cal{D}} \alpha_i \tilde
\varphi_{\|} (\alpha_i) e^{iq \cdot x(\alpha_1 +v \alpha_3)} \hskip 3 pt
.
\end{eqnarray}
Here the operator $\tilde G_{\alpha \beta}$  is the dual of
$G_{\alpha \beta}$, $\tilde G_{\alpha \beta}= {1\over 2}
\epsilon_{\alpha \beta \delta \rho} G^{\delta \rho} $,
${\cal{D}}\alpha_i$ is defined as ${\cal{D}} \alpha_i =d \alpha_1
d \alpha_2 d \alpha_3 \delta(1-\alpha_1 -\alpha_2 -\alpha_3)$ and
$\Phi(\alpha_1,\alpha_2,\alpha_3)=\varphi_{\bot}+\varphi_{\|}-\tilde
\varphi_{\bot}-\tilde \varphi_{\|}$.

The twist-3 and twist-4 light-cone distribution amplitudes can be
parameterized as
\begin{eqnarray}
\varphi_p(u,\mu)&=&1+\left(30\eta_3-\frac{5}{2}\rho^2\right)C_2^{\frac{1}{2}}(2u-1)\nonumber \\
&+&\left(-3\eta_3\omega_3-\frac{27}{20}\rho^2-\frac{81}{10}\rho^2\tilde{a}_2\right)C_4^{\frac{1}{2}}(2u-1) \nonumber \\
\varphi_\sigma(u,\mu)&=&6u(1-u)\left(1
+\left(5\eta_3-\frac{1}{2}\eta_3\omega_3-\frac{7}{20}\rho^2-\frac{3}{5}\rho^2\tilde{a}_2\right)C_2^{\frac{3}{2}}(2u-1)\right), \nonumber \\
\phi_{3K}(\alpha_i,\mu) &=& 360 \alpha_1 \alpha_2 \alpha_3^2 \left (1 + a(\mu) \frac{1}{2} ( 7 \alpha_3 - 3) + b(\mu)
(2 - 4 \alpha_1 \alpha_2 - 8 \alpha_3 (1 - \alpha_3)) \right . \nonumber \\
& &  + \left . c(\mu) ( 3 \alpha_1 \alpha_2 - 2 \alpha_3 + 3 \alpha_3^2) \right ) \, ,
 \nonumber\\
\phi_{\perp}(\alpha_i,\mu) &=& 30 \delta^2(\mu)(\alpha_1-\alpha_2)\alpha_3^2\left [ \frac{1}{3} +
2 \epsilon (\mu) (1 - 2 \alpha_3) \right ]  \, ,
\nonumber \\
\phi_{||}(\alpha_i,\mu) &=& 120 \delta^2(\mu) \epsilon (\mu)  (\alpha_1-\alpha_2) \alpha_1 \alpha_2 \alpha_3  \, ,
\nonumber\\
\tilde{\phi}_{\perp}(\alpha_i,\mu) &=&
30 \delta^2(\mu) \alpha_3^2 ( 1 - \alpha_3) \left [ \frac{1}{3} + 2 \epsilon (\mu)  (1 - 2 \alpha_3) \right ] \, ,
 \nonumber\\
\tilde{\phi}_{||}(\alpha_i,\mu) &=& -120 \delta^2(\mu) \alpha_1 \alpha_2 \alpha_3 \left [ \frac{1}{3} +
\epsilon (\mu) (1 - 3 \alpha_3) \right ] \, ,
 \
\end{eqnarray}
where  $ C_2^{\frac{1}{2}}$, $ C_4^{\frac{1}{2}}$
 and $ C_2^{\frac{3}{2}}$ are Gegenbauer polynomials.
The parameters in the light-cone distribution amplitudes can be estimated
from the QCD sum rules approach \cite{Kho1,CZ,BF90}. In practical
manipulation, we choose $a=-2.88$, $b=0.0$, $c=0.0$,
$\delta^2=0.2GeV^2$ and $\epsilon=0.5$ at $\mu=1GeV$ .
Furthermore, the updated values for $\eta_3$ , $ \omega_3 $, $
\rho$ and $ \tilde a_2$ are taken as $\tilde a_2=0.2$,
$\eta_3=0.015$,  $\omega_3=-3$ at the scale $\mu \simeq 1$ GeV and
the parameter $\rho^2={m_s^2\over M_K^2}$
 \cite{Belyaev:1994zk,Ball:1998je}.

Now we perform the Borel transformation with respect to  the
variables $Q_1^2=-p^2$ and  $Q_2^2=-(p+q)^2$ for the correlation
functions in Eqs.(8-9) and obtain the analytical expressions for the
invariant functions in the hadronic representation,
\begin{eqnarray}
B_{M_2^2}B_{M_1^2}T^{f_0}_p(M_1^2,M_2^2)&=&if_{K} g_{f_0K^+K^-}f_{f_0}M_{f_0}  \frac{1}{M^2_1M^2_2} e^{-M^2_K/M_1^2}e^{-M^2_{f_0}/M_2^2}\nonumber\\
&&+\frac{i}{M^2_1M^2_2}\int^\infty_{s_{0}}ds\int^\infty_{s'_{0}}ds'~\rho^{cont}(s,s')e^{-s/M_1^2}e^{-s'/M_2^2}~~, \\
B_{M_2^2}B_{M_1^2}T^{a_0}_p(M_1^2,M_2^2)&=&if_{K} g_{a_0K^+K^-} f_{a_0}M_{a_0}  \frac{1}{M^2_1M^2_2}e^{-M^2_K/M_1^2}e^{-M^2_{a_0}/M_2^2}\nonumber\\
&&+\frac{i}{M^2_1M^2_2}\int^\infty_{s_{0}}ds\int^\infty_{s'_{0}}ds'~\rho^{cont}(s,s')e^{-s/M_1^2}e^{-s'/M_2^2}~~,
\end{eqnarray}
here we have not shown  the cross terms explicitly for simplicity.    In order to match the duality regions below the thresholds $s_0$
and $s_0'$, we can express the correlation functions at the level of
quark-gluon  degrees of freedom into the following form,
\begin{eqnarray}
T^{f_0}_p(p^2,(p+q)^2)= i\int ds ds^\prime {\rho^{f_0}_{quark}(s,s^\prime) \over
(s-p^2) [s^\prime-(p+q)^2]} \, , \\
T^{a_0}_p(p^2,(p+q)^2)= i\int ds ds^\prime {\rho^{a_0}_{quark}(s,s^\prime) \over
(s-p^2) [s^\prime-(p+q)^2]} \, .
\end{eqnarray}
Then it is a straightforward   procedure to perform the Borel
transformation with respect to the variables $Q_1=-p^2$ and
$Q_2^2=-(p+q)^2$, however, the analytical expressions for the
spectral densities $\rho^{f_0}_{quark}(s,s')$, $\rho^{a_0}_{quark}(s,s')$ are hard to obtain, we
have to take some approximations,  as the contributions
 from the higher twist terms  are  suppressed by more powers of  $\frac{1}{-p^2}$ or $\frac{1}{-(p+q)^2}$
and the continuum subtractions will not affect the results
remarkably, here we will use the expressions in Eqs.(12-13) for the three-particle (quark-antiquark-gluon)
twist-3 and twist-4  terms.
As for the terms involving $\varphi_p$ and $\varphi_\sigma$ , we
preform the same type trick as Refs.\cite{Belyaev:1994zk,Kim} and
expand the amplitudes $\varphi_p(u)$ and $\varphi_\sigma(u)$ in
terms of polynomials of $1-u$,
\begin{eqnarray}
\varphi_p(u)+{d\varphi_\sigma(u)\over
6du}=\sum_{k=0}^N b_k (1-u)^k,
\end{eqnarray}
then the variable $u$ is changed into $s$ , $s'$ and the spectral
densities are obtained.

After straightforward but cumbersome calculations, we obtain the final expressions for the
Borel transformed correlation functions at the level of quark-gluon degrees of freedom,
\begin{eqnarray}
B_{M_2^2}B_{M_1^2}T^{f_0}_p &=& \frac{sin\theta}{\sqrt{2}} \frac{i}{M_1^2M_2^2}
e^{-\frac{u_0(1-u_0)M_K^2}{M^2}}
\left\{ {f_K M^2 M_K^2 \over m_s} \, \, \sum_{k=0}^N b_k ({M^2 \over M_1^2})^k
\left[1- e^{-\frac{s_0}{M^2}} \sum_{i=0}^k {\left(\frac{s_0}{M^2}\right)^i \over i!} \right]
  \right.\nonumber \\
&& \left.+f_{3K} M_K^2  \int_0^{u_0} d \alpha_1
\int_{u_0-\alpha_1}^{1-\alpha_1} {d \alpha_3 \over \alpha_3}
\varphi_{3K}(\alpha_1,1-\alpha_1-\alpha_3,\alpha_3)
\left(2{u_0-\alpha_1 \over \alpha_3}-3 \right)  \right\}   \nonumber\\
&&+cos\theta \frac{i}{M_1^2M_2^2}e^{-\frac{m_s^2+u_0(1-u_0)M_K^2}{M^2}}
\left\{ {f_K M^2 M_K^2 \over m_s} \, \, \sum_{k=0}^N b_k ({M^2 \over M_1^2})^k
\left[1- e^{-\frac{s_0-m_s^2}{M^2}} \sum_{i=0}^k {\left(\frac{s_0-m_s^2}{M^2}\right)^i \over i!} \right]
 \right. \nonumber \\
&&\left.-2 f_K m_s  \, g_2(u_0)+f_{3K} M_K^2  \int_0^{u_0} d \alpha_1
\int_{u_0-\alpha_1}^{1-\alpha_1} {d \alpha_3 \over \alpha_3}
\varphi_{3K}(\alpha_1,1-\alpha_1-\alpha_3,\alpha_3)
\left(2{u_0-\alpha_1 \over \alpha_3}-3 \right) \right.\nonumber \\
&&\left.- { 2f_K m_s M_K^2\over M^2} (1-u_0) \int_{1-u_0}^1 {d \alpha_3 \over
\alpha_3^2}  \int_0^{\alpha_3}d \beta \int_0^{1-\beta}d \alpha \Phi(\alpha,1-\alpha-\beta,\beta)  \right. \nonumber \\
&&\left.+ {2f_Km_s M_K^2\over M^2}\left[ \int_0^{1-u_0} {d \alpha_3 \over
\alpha_3} \int_{u_0-\alpha_3}^{u_0} d \alpha_1 \int_0^{\alpha_1}d\alpha+ \int_{1-u_0}^1 {d \alpha_3 \over \alpha_3}
\int_{u_0-\alpha_3}^{1-\alpha_3} d \alpha_1 \int_0^{\alpha_1}d\alpha
\right]\Phi(\alpha,1-\alpha-\alpha_1,\alpha_1)  \right\} \nonumber \,, \\
\end{eqnarray}
\begin{eqnarray}
B_{M_2^2}B_{M_1^2}T^{a_0}_p &=& \frac{sin\varphi}{\sqrt{2}} \frac{i}{M_1^2M_2^2}
e^{-\frac{u_0(1-u_0)M_K^2}{M^2}}
\left\{ {f_K M^2 M_K^2 \over m_s} \, \, \sum_{k=0}^N b_k ({M^2 \over M_1^2})^k
\left[1- e^{-\frac{s_0}{M^2}} \sum_{i=0}^k {\left(\frac{s_0}{M^2}\right)^i \over i!} \right]
  \right.\nonumber \\
&& \left.+f_{3K} M_K^2  \int_0^{u_0} d \alpha_1
\int_{u_0-\alpha_1}^{1-\alpha_1} {d \alpha_3 \over \alpha_3}
\varphi_{3K}(\alpha_1,1-\alpha_1-\alpha_3,\alpha_3)
\left(2{u_0-\alpha_1 \over \alpha_3}-3 \right)  \right\}   \nonumber\\
&&+cos\varphi \frac{i}{M_1^2M_2^2}e^{-\frac{m_s^2+u_0(1-u_0)M_K^2}{M^2}}
\left\{ {f_K M^2 M_K^2 \over m_s} \, \, \sum_{k=0}^N b_k ({M^2 \over M_1^2})^k
\left[1- e^{-\frac{s_0-m_s^2}{M^2}} \sum_{i=0}^k {\left(\frac{s_0-m_s^2}{M^2}\right)^i \over i!} \right]
 \right. \nonumber \\
&&\left.-2 f_K m_s  \, g_2(u_0)+f_{3K} M_K^2  \int_0^{u_0} d \alpha_1
\int_{u_0-\alpha_1}^{1-\alpha_1} {d \alpha_3 \over \alpha_3}
\varphi_{3K}(\alpha_1,1-\alpha_1-\alpha_3,\alpha_3)
\left(2{u_0-\alpha_1 \over \alpha_3}-3 \right) \right.\nonumber \\
&&\left.- { 2f_K m_s M_K^2\over M^2} (1-u_0) \int_{1-u_0}^1 {d \alpha_3 \over
\alpha_3^2}  \int_0^{\alpha_3}d \beta \int_0^{1-\beta}d \alpha \Phi(\alpha,1-\alpha-\beta,\beta)  \right. \nonumber \\
&&\left.+ {2f_Km_s M_K^2\over M^2}\left[ \int_0^{1-u_0} {d \alpha_3 \over
\alpha_3} \int_{u_0-\alpha_3}^{u_0} d \alpha_1 \int_0^{\alpha_1}d\alpha+ \int_{1-u_0}^1 {d \alpha_3 \over \alpha_3}
\int_{u_0-\alpha_3}^{1-\alpha_3} d \alpha_1 \int_0^{\alpha_1}d\alpha
\right]\Phi(\alpha,1-\alpha-\alpha_1,\alpha_1)  \right\} \nonumber \, . \\
\end{eqnarray}
In deriving the above expressions for
$\varphi_p(u)+{d\varphi_\sigma(u)\over 6du}$, we have neglected
the terms $\sim M_K^4$ . Here $u_0=\frac{M_1^2}{M_1^2+M_2^2}$ and
$M^2=\frac{M_1^2M_2^2}{M_1^2+M_2^2}$.

The matching between Eqs.(16-17) and Eqs.(21-22) below the
thresholds  $s_0$, $s_0'$ is straightforward  and we can obtain
the analytical  expressions for the strong coupling constants $g_{f_0K^+K^-}$ and $g_{a_0K^+K^-}$,
\begin{eqnarray}
g_{f_0K^+K^-}&=& \frac{sin\theta}{\sqrt{2}} \frac{1}{f_K f_{f_0}M_{f_0}}
e^{\frac{M^2_{f_0}}{M_2^2}+\frac{M^2_K}{M_1^2}-\frac{u_0(1-u_0)M_K^2}{M^2}}
\left\{ {f_K M^2 M_K^2 \over m_s} \, \, \sum_{k=0}^N b_k ({M^2 \over M_1^2})^k
\left[1- e^{-\frac{s_0}{M^2}} \sum_{i=0}^k {\left(\frac{s_0}{M^2}\right)^i \over i!} \right]
  \right.\nonumber \\
&& \left.+f_{3K} M_K^2  \int_0^{u_0} d \alpha_1
\int_{u_0-\alpha_1}^{1-\alpha_1} {d \alpha_3 \over \alpha_3}
\varphi_{3K}(\alpha_1,1-\alpha_1-\alpha_3,\alpha_3)
\left(2{u_0-\alpha_1 \over \alpha_3}-3 \right)  \right\}   \nonumber\\
&&+cos\theta \frac{1}{f_K f_{f_0}M_{f_0}}e^{\frac{M^2_{f_0}}{M_2^2}+\frac{M^2_K}{M_1^2}-\frac{m_s^2+u_0(1-u_0)M_K^2}{M^2}}
\left\{ {f_K M^2 M_K^2 \over m_s} \, \, \sum_{k=0}^N b_k ({M^2 \over M_1^2})^k
\left[1- e^{-\frac{s_0-m_s^2}{M^2}} \sum_{i=0}^k {\left(\frac{s_0-m_s^2}{M^2}\right)^i \over i!} \right]
 \right. \nonumber \\
&&\left.-2 f_K m_s  \, g_2(u_0)+f_{3K} M_K^2  \int_0^{u_0} d \alpha_1
\int_{u_0-\alpha_1}^{1-\alpha_1} {d \alpha_3 \over \alpha_3}
\varphi_{3K}(\alpha_1,1-\alpha_1-\alpha_3,\alpha_3)
\left(2{u_0-\alpha_1 \over \alpha_3}-3 \right) \right.\nonumber \\
&&\left.- { 2f_K m_s M_K^2\over M^2} (1-u_0) \int_{1-u_0}^1 {d \alpha_3 \over
\alpha_3^2}  \int_0^{\alpha_3}d \beta \int_0^{1-\beta}d \alpha \Phi(\alpha,1-\alpha-\beta,\beta)  \right. \nonumber \\
&&\left.+ {2f_Km_s M_K^2\over M^2}\left[ \int_0^{1-u_0} {d \alpha_3 \over
\alpha_3} \int_{u_0-\alpha_3}^{u_0} d \alpha_1 \int_0^{\alpha_1}d\alpha+ \int_{1-u_0}^1 {d \alpha_3 \over \alpha_3}
\int_{u_0-\alpha_3}^{1-\alpha_3} d \alpha_1 \int_0^{\alpha_1}d\alpha
\right]\Phi(\alpha,1-\alpha-\alpha_1,\alpha_1)  \right\} \nonumber   \\
&=& sin\theta g^{\bar{n}n}_{f_0K^+K^-}+cos\theta g^{\bar{s}s}_{f_0K^+K^-};
\end{eqnarray}
\begin{eqnarray}
g_{a_0K^+K^-} &=& \frac{sin\varphi}{\sqrt{2}} \frac{1}{f_K f_{a_0}M_{a_0}}
e^{\frac{M^2_{a_0}}{M_2^2}+\frac{M^2_K}{M_1^2}-\frac{u_0(1-u_0)M_K^2}{M^2}}
\left\{ {f_K M^2 M_K^2 \over m_s} \, \, \sum_{k=0}^N b_k ({M^2 \over M_1^2})^k
\left[1- e^{-\frac{s_0}{M^2}} \sum_{i=0}^k {\left(\frac{s_0}{M^2}\right)^i \over i!} \right]
  \right.\nonumber \\
&& \left.+f_{3K} M_K^2  \int_0^{u_0} d \alpha_1
\int_{u_0-\alpha_1}^{1-\alpha_1} {d \alpha_3 \over \alpha_3}
\varphi_{3K}(\alpha_1,1-\alpha_1-\alpha_3,\alpha_3)
\left(2{u_0-\alpha_1 \over \alpha_3}-3 \right)  \right\}   \nonumber\\
&&+cos\varphi \frac{1}{f_K f_{a_0}M_{a_0}}e^{\frac{M^2_{a_0}}{M_2^2}+\frac{M^2_K}{M_1^2}-\frac{m_s^2+u_0(1-u_0)M_K^2}{M^2}}
\left\{ {f_K M^2 M_K^2 \over m_s} \, \, \sum_{k=0}^N b_k ({M^2 \over M_1^2})^k
\left[1- e^{-\frac{s_0-m_s^2}{M^2}} \sum_{i=0}^k {\left(\frac{s_0-m_s^2}{M^2}\right)^i \over i!} \right]
 \right. \nonumber \\
&&\left.-2 f_K m_s  \, g_2(u_0)+f_{3K} M_K^2  \int_0^{u_0} d \alpha_1
\int_{u_0-\alpha_1}^{1-\alpha_1} {d \alpha_3 \over \alpha_3}
\varphi_{3K}(\alpha_1,1-\alpha_1-\alpha_3,\alpha_3)
\left(2{u_0-\alpha_1 \over \alpha_3}-3 \right) \right.\nonumber \\
&&\left.- { 2f_K m_s M_K^2\over M^2} (1-u_0) \int_{1-u_0}^1 {d \alpha_3 \over
\alpha_3^2}  \int_0^{\alpha_3}d \beta \int_0^{1-\beta}d \alpha \Phi(\alpha,1-\alpha-\beta,\beta)  \right. \nonumber \\
&&\left.+ {2f_Km_s M_K^2\over M^2}\left[ \int_0^{1-u_0} {d \alpha_3 \over
\alpha_3} \int_{u_0-\alpha_3}^{u_0} d \alpha_1 \int_0^{\alpha_1}d\alpha+ \int_{1-u_0}^1 {d \alpha_3 \over \alpha_3}
\int_{u_0-\alpha_3}^{1-\alpha_3} d \alpha_1 \int_0^{\alpha_1}d\alpha
\right]\Phi(\alpha,1-\alpha-\alpha_1,\alpha_1)  \right\} \nonumber   \\
&=&sin\varphi g^{\bar{n}n}_{a_0K^+K^-}+cos\varphi g^{\bar{s}s}_{a_0K^+K^-}.
\end{eqnarray}
The values of the parameters $ f_{f_0} $
and $f_{a_0} $ can be determined from the conventional QCD sum rules approach
with the following two two-point correlation functions,
\begin{eqnarray}
T_{f_0}&=&i \int d^4 x e^{ipx}\langle |T[J_{f_0}(x)J_{f_0}(0)]|\rangle \nonumber  \\
&=&sin^2\theta i \int d^4 x e^{ipx}\langle |T[\frac{\bar{u}u+\bar{d}d}{\sqrt{2}}(x)\frac{\bar{u}u+\bar{d}d}{\sqrt{2}}(0)]|\rangle
+cos^2\theta i \int d^4 x e^{ipx}\langle |T[ \bar{s}s  (x)\bar{s}s(0) ]|\rangle \, ,\\
T_{a_0}&=&i \int d^4 x e^{ipx}\langle |T[J_{a_0}(x)J_{a_0}(0)]|\rangle \nonumber  \\
&=&sin^2\varphi i \int d^4 x e^{ipx}\langle |T[\frac{\bar{u}u-\bar{d}d}{\sqrt{2}}(x)\frac{\bar{u}u-\bar{d}d}{\sqrt{2}}(0)]|\rangle
+cos^2\varphi i \int d^4 x e^{ipx}\langle |T[ \bar{s}s  (x)\bar{s}s(0) ]|\rangle \,.
\end{eqnarray}
The operator product expansion near the $x\sim 0$ in perturbative QCD is straightforward
  and we will not write down the
detailed routine for simplicity. The final expressions for the $T_{f_0}$ and $T_{a_0}$ at the level of quark-gluon degrees of freedom
can be written as
\begin{eqnarray}
T_{f_0}&=&sin^2\theta A(p^2)+cos^2\theta B(p^2) \, , \nonumber  \\
T_{a_0}&=&sin^2\varphi C(p^2)+cos^2\varphi D(p^2)\, ,
\end{eqnarray}
here $A,B,C,D$ are formal notations.
 To obtain the hadronic representation, we can insert a complete series of intermediate states with
the same quantum numbers as the interpolating currents $ J_{f_0}$ and $J_{a_0}$ into the correlation functions
in Eqs.(25-26), then
 isolate the ground state
contributions from the $f_0(980)$ and $a_0(980)$ mesons,
\begin{eqnarray}
T_{f_0}&=&\langle |J_{f_0}(0)| f_0 (p)\rangle \frac{1}{M_{f_0}^2-p^2}\langle f_0(p)|J_{f_0}(0)|\rangle+\cdots, \nonumber \\
&=&\frac{f_{f_0}^2   M^2_{f_0}}{M_{f_0}^2-p^2}+\cdots \ \  ,\nonumber \\
&=&sin^2\theta\langle |\frac{\bar{u}u+\bar{d}d}{\sqrt{2}}(0)| f_0 (p)\rangle \frac{1}
{M_{f_0}^2-p^2}\langle f_0(p)|\frac{\bar{u}u+\bar{d}d}{\sqrt{2}}(0)|\rangle \nonumber \\
&&+cos^2\theta\langle |\bar{s}s(0)| f_0 (p)\rangle \frac{1}{M_{f_0}^2-p^2}\langle f_0(p)|\bar{s}s(0)|\rangle+\cdots, \nonumber \\
&=&sin^2\theta\frac{f_{\bar{n}nf_0}^2   M^2_{f_0}}{M_{f_0}^2-p^2} +cos^2\theta\frac{f_{\bar{s}sf_0}^2
M^2_{f_0}}{M_{f_0}^2-p^2}+\cdots \ \  ,\nonumber \\
 T_{a_0}&=&\langle |J_{a_0}(0)| a_0 (p)\rangle \frac{1}{M_{a_0}^2-p^2}\langle a_0(p)|J_{a_0}(0)|\rangle+\cdots, \nonumber \\
&=&\frac{f_{a_0}^2   M^2_{a_0}}{M_{a_0}^2-p^2}+\cdots \ \ , \nonumber \\
&=&sin^2\varphi\langle |\frac{\bar{u}u-\bar{d}d}{\sqrt{2}}(0)| a_0 (p)\rangle \frac{1}
{M_{a_0}^2-p^2}\langle a_0(p)|\frac{\bar{u}u-\bar{d}d}{\sqrt{2}}(0)|\rangle \nonumber \\
&&+cos^2\varphi\langle |\bar{s}s(0)| a_0 (p)\rangle \frac{1}{M_{a_0}^2-p^2}\langle a_0(p)|\bar{s}s(0)|\rangle+\cdots, \nonumber \\
&=&sin^2\varphi\frac{f_{\bar{n}na_0}^2   M^2_{a_0}}{M_{a_0}^2-p^2} +cos^2
\varphi\frac{f_{\bar{s}sa_0}^2   M^2_{a_0}}{M_{a_0}^2-p^2}+\cdots \ \ .
\end{eqnarray}
Here we have used the following definitions,
\begin{eqnarray}
\langle\mid \frac{\bar{u}u+\bar{d}d}{\sqrt{2}}(0)\mid f_0(p)\rangle &=&f_{\bar{n}nf_0}M_{f_0}, \nonumber \\
\langle\mid \frac{\bar{u}u-\bar{d}d}{\sqrt{2}}(0)\mid a_0(p)\rangle &=&f_{\bar{n}na_0}M_{a_0}, \nonumber \\
\langle\mid  \bar{s}s (0)\mid f_0(p)\rangle &=&f_{\bar{s}sf_0}M_{f_0}, \nonumber \\
\langle\mid  \bar{s}s (0)\mid a_0(p)\rangle &=&f_{\bar{s}sa_0}M_{a_0}.
\end{eqnarray}
After performing the standard manipulations of the quark-hadron
duality ( i.e. matching Eq.(27) to Eq.(28) ) and Borel transformations,  we can  equate  the coefficients of the
$sin^2\theta , cos^2\theta , sin^2\varphi $ and $cos^2\varphi $, respectively. Finally  we obtain the decay constants  (coupling constants),
\begin{eqnarray}
f_{\bar{n}nf_0}  =f_{\bar{n}na_0}=214\pm10MeV \,,  \ \ \ \ f_{\bar{s}sf_0} =f_{\bar{s}sa_0}= 180\pm10MeV  \, , \nonumber \\
f_{f_0}=\sqrt{sin^2\theta f^2_{\bar{n}nf_0}+cos^2\theta f^2_{\bar{s}sf_0}} \,, \ \ \ f_{a_0}=\sqrt{sin^2\varphi f^2_{\bar{n}na_0}+cos^2\varphi f^2_{\bar{s}sa_0}} \,.
\end{eqnarray}
The existing values for the mixing angle $\theta$ differ from each other greatly,
the analysis of the $J/\psi$ decays indicates $\theta=(34\pm 6) ^\circ$ or $\theta=(146\pm 6) ^\circ$ \cite{ChengHY}
while the analysis of the $D^+_s$ decays
  $D^+_s\rightarrow f_0(980)\pi^+$ and $D^+_s\rightarrow \phi\pi^+$ indicates $35^\circ  \leq -\theta \leq 55^\circ$ \cite{Anisovich03}.
If the value $\theta=(34\pm 6) ^\circ$ is taken, we can obtain $f_{f_0}=191\pm13 MeV$.
The values for the decay constants $f_{\bar{n}nf_0}(f_{\bar{n}na_0})$ and  $f_{\bar{s}sf_0}(f_{\bar{s}sa_0})$ are close to each other,
the variations of $\theta$ and $\varphi$ will not lead to significant changes for the net decay constants $f_{f_0}$ and $f_{a_0}$, in the following,
we take the values
$f_{f_0}=f_{a_0}=191\pm13 MeV$ for simplicity. The  simplification will obviously introduce some imprecision,
however, the strong coupling constants $g_{f_0K^+K^-}(g_{a_0K^+K^-})\sim \frac{1}{f_{f_0}}(\frac{1}{f_{a_0}})$,
the  final results will not be remarkably  affected.

To obtain the above values in Eq.(30) for the two-point correlation functions in  Eqs.(25-26), the vacuum condensates are taken as
$\langle \bar{s}s \rangle=0.8\langle \bar{u}u \rangle$, $\langle \bar{u}u \rangle=\langle \bar{d}d \rangle=(-240\pm 10 MeV)^3$,
$\langle \bar{s}\sigma \cdot G s \rangle=(0.8\pm 0.1)\langle \bar{s}s \rangle$,
 $\langle \bar{u}\sigma \cdot G u \rangle=(0.8\pm 0.1)\langle \bar{u}u \rangle$,
$\langle \bar{d}\sigma \cdot G d \rangle=(0.8\pm 0.1)\langle \bar{d}d \rangle$ , $M_{f_0}=M_{a_0}=980MeV$.
 The threshold parameter $s_0$ is chosen to  vary between $1.6-1.7GeV^2$ to avoid possible pollutions from
  higher resonances and continuum states. In the region $1.2-2.0GeV^2$, the sum rules are almost independent of
 the Borel parameter $M^2$.

Now we return to the values of  the strong coupling constants
$g_{f_0K^+K^-}$, $g_{a_0K^+K^-}$, and choose the parameters as $m_s=150MeV$,
$f_{3K}=f_{3\pi}=0.0035GeV^2$ at about $\mu=1 GeV$,
$f_K=0.160GeV$ and $M_K=498MeV$ \cite{Kho1,CZ,BF90}. The duality
thresholds in Eqs.(23-24) are taken as $s_0=1.0-1.1GeV^2$ determined from two point K meson QCD sum rules
 to avoid possible  pollutions from the higher resonances and
continuum states.  The Borel parameters are chosen as $0.8
\le M_1^2 \le 1.6$ GeV$^2$ and $2.0\le M_2^2 \le 4.5$ GeV$^2$, in those  regions, the values for
the strong coupling constants $g_{f_0 K^+ K^-}$ and $g_{f_0 K^+ K^-}$ are rather stable.
Finally the numerical results for the strong coupling constants are
obtained,
\begin{eqnarray}
6.1 \le g^{\bar{s}s}_{f_0 K^+ K^-}(g^{\bar{s}s}_{a_0 K^+ K^-}) \le 7.5 GeV   ; \\
4.4 \le g^{\bar{n}n}_{f_0 K^+ K^-}(g^{\bar{n}n}_{a_0 K^+ K^-}) \le 5.5 GeV  .
\end{eqnarray}
\begin{eqnarray}
\theta=(34\pm 6)^\circ\cite{ChengHY}, \ \ \ g_{f_0 K^+ K^-}=7.4\sim9.3 \ \ ; && \ \ \
\theta=(146\pm 6)^\circ, \ \ \ g_{f_0 K^+ K^-}=-4.0\sim-1.8\,;\nonumber\\
\theta=(-35\sim-55)^\circ\cite{Anisovich03}, \ \ \ g_{f_0 K^+ K^-}=-0.2\sim3.0\ \ ; && \ \ \
\theta=(-15\sim-35)^\circ, \ \ \ g_{f_0 K^+ K^-}=3.0\sim5.8\,;\nonumber\\
\varphi=(-30\sim-40)^\circ, \ \ \ g_{a_0 K^+ K^-}=1.8\sim3.7 \ \ ; && \ \ \
\varphi=80^\circ, \ \ \ g_{a_0 K^+ K^-}=5.4\sim6.8 \, ; \nonumber\\
\varphi=90^\circ, \ \ \ g_{a_0 K^+ K^-}=4.4\sim5.5\ \ ; && \ \ \
\varphi=100^\circ, \ \ \ g_{a_0 K^+ K^-}=3.3\sim4.1\,.
\end{eqnarray}
From the above numerical
results, in spite of the constituent structure differences between the
$f_0(980)$ and $a_0(980)$ mesons, we can see that the strong
couplings to the S-wave $K^+K^-$ state through the $s\bar{s}$ components are larger than
the corresponding ones through the $n\bar{n}$ components, $g_{f_0
K^+ K^-}^{\bar{s}s}\approx \sqrt{2}g_{f_0 K^+ K^-}^{\bar{n}n}$ and
$g_{a_0 K^+ K^-}^{\bar{s}s}\approx \sqrt{2} g_{a_0 K^+
K^-}^{\bar{n}n}$. Due to the special Dirac structures of the interpolating currents $J_{f_0}$ and $J_{a_0}$,
the values of the strong $K^+K^-$ couplings components of the $a_0(980)$ meson are about the same as the corresponding ones for
the $f_0(980)$ meson, $g^{\bar{s}s}_{f_0 K^+ K^-} \approx g^{\bar{s}s}_{a_0 K^+ K^-}$,
$g^{\bar{n}n}_{f_0 K^+ K^-}\approx g^{\bar{n}n}_{a_0 K^+ K^-} $.
 Furthermore, the strong coupling constants $g_{f_0K^+K^-}$ and $g_{a_0K^+K^-}$
are nearly linear functions of the $cos\theta$, $sin\theta$, $cos\varphi$ and $sin\varphi$ (see Eqs.(23-24)),
 the variations with respect
to the parameters $\theta$ and $\varphi$ can  change their values significantly i.e. they are sensitive to the mixing angles.

In the following, we list the experimental data for the values of
the strong coupling constants $g_{f_0K^+K^-}$ and $g_{a_0K^+K^-}$.
  $g_{f_0 K^+ K^-} =4.0\pm 0.2 GeV$ by KLOE Collaboration see Ref.\cite{Aloisio:2002bt},
  $g_{f_0 K^+ K^-} =4.3\pm 0.5 GeV$ by CMD-2 Collaboration see Ref.\cite{Akhmetshin:1999di},
    $g_{f_0 K^+ K^-} =5.6\pm 0.8$ by  SND Collaboration  see Ref.\cite{Achasov:2000ym},
                $g_{f_0 K^+ K^-} =2.2\pm 0.2$ by WA102 Collaboration see  Ref.\cite{Barberis:1999cq},
  $g_{f_0 K^+ K^-} =0.5\pm 0.6$ by  E791  Collaboration   see  Ref. \cite{Gobel:2000es},
 $g_{a_0K^+K^-}=2.3\pm 0.7GeV$ by KLOE collaboration see Ref. \cite{KLOEa0}
 and
$g_{a_0K^+K^-}=2.63^{+1.84}_{-1.28}GeV$ by analysis of KLOE collaboration data see Ref. \cite{Achasova0}.
While the theoretical values are $g_{f_0 K^+ K^-}=2.24  GeV $ by
linear sigma model see Ref.\cite{Napsuciale:1998ip}, $g_{f_0 K^+
K^-}=3.68\pm0.13 GeV$ and $g_{a_0 K^+ K^-}=5.50\pm0.11 GeV$ by
unitary Chiral perturbation theory see Ref.\cite{Oller2003}.

Comparing with all the  controversial values, we can not reach  a
general consensus on the strong coupling constants  $g_{f_0 K^+
K^-}$ and $g_{a_0 K^+ K^-}$. If we take the mixing angle $\theta=-15^\circ\sim-35^\circ$ for the $f_0(980)$ meson,
the value of the strong coupling constant $g_{f_0K^+K^-}$ is  $g_{f_0K^+K^-}=3.0\sim5.8$, which is considerably compatible with the existing experimental data. For the
$a_0(980)$ meson, no conclusion can be made from the existing values for the mixing angle $\varphi$.
 Precise determination of those values call for more accurate measures and original theoretical
approaches. Despite whatever the mixing angles $\theta$, $\varphi$ may be,
we observe that the strong couplings through both the $n\bar{n}$ and $s\bar{s}$ components are remarkably large. This  fact
obviously supports the hadronic dressing mechanism, the $f_0(980)$ and $a_0(980)$ mesons can be
taken  as have small $q\bar{q}$ kernels of typical meson size with large virtual S-wave $K\bar{K}$ cloud.

\section{Conclusions}

In this article, with the assumption of explicit isospin violation arising from the $f_0(980)-a_0(980)$ mixing,
we take the point of view that the  $f_0(980)$ and
$a_0(980)$   mesons have both strange and non-strange $q\bar{q}$ components and
evaluate the strong coupling constants $g_{f_0 K^+ K^-}$ and
$g_{a_0 K^+ K^-}$  within the framework of the light-cone QCD sum rules  approach.
Taking into account the controversial values emerge from different
experimental and theoretical determinations, we can not reach a
general consensus. Our observation about the large  scalar-$KK$
coupling constants $g^{\bar{n}n}_{f_0 K^+ K^-}$,
$g^{\bar{s}s}_{f_0 K^+ K^-}$, $g^{\bar{n}n}_{a_0 K^+ K^-}$ and $g^{\bar{s}s}_{a_0 K^+ K^-}$
based on the light-cone QCD sum rules approach will support the
hadronic dressing mechanism, furthermore, in spite of the constituent  structure
differences  between the $f_0(980)$ and $a_0(980)$ mesons, the
strange components have  larger strong coupling constants with the
$K^+K^-$  state  than the corresponding non-strange ones, $g_{f_0 K^+
K^-}^{\bar{s}s}\approx \sqrt{2}g_{f_0 K^+ K^-}^{\bar{n}n}$ and
$g_{a_0 K^+ K^-}^{\bar{s}s}\approx \sqrt{2} g_{a_0 K^+
K^-}^{\bar{n}n}$.

{\large Note Added,}

The interest in the nature of the light scalar $f_0(980)$ and $a_0(980)$ mesons and their mixing
was renewed recently. There has been a number of articles attempting  to elucidate those elusive mesons since
we have finished our article,
for example, Ref. \cite{2004}.


\begin{thebibliography}{99}
\bibitem{Godfray} S. Godfray and J. Napolitano, Rev. Mod. Phys. {\bf 71 } (1999) 1411.

\bibitem{Close2002} F. E. Close and N. A. Tornqvist, J. Phys.  {\bf G28} (2002) R249.

\bibitem{JaffeAchasov4q} R. L . Jaffe and K. Johnson, Phys. Lett.  {\bf B60} (1976) 201;
R. L. Jaffe, Phys. Rev. {\bf D15} (1977) 267, 281; {\bf D17} (1978) 1444;
  N. N. Achasov and V. N. Ivanchenko, Nucl. Phys.  {\bf B315} (1989) 465;
N. N. Achasov and V. V. Gubin, Phys. Rev.  {\bf D56} (1997) 4084;
Phys. Rev.  {\bf D63} (2001) 094007.

\bibitem{IsgurKK} J. Weinstein and N. Isgur, Phys. Rev. Lett. {\bf 48} (1982) 659;
Phys. Rev.  {\bf D27} (1983) 588; Phys. Rev. {\bf D41} (1990) 2236 ;
F. E. Close, N. Isgur and S. Kumana, Nucl. Phys. {\bf B389} (1993) 513;
 R. Kaminski, L. Lesniak and J. P. Maillet, Phys. Rev. {\bf D50} (1994) 3145;
N. N. Achasov, V. V. Gubin and V. I. Shevchenko, Phys. Rev. {\bf D56} (1997) 203;
Yu. S. Surovtsev, D. Krupa and M. Nagy, hep-ph/0311195.


\bibitem{HDress} N. A. Tornqvist, Z. Phys. {\bf C68} (1995) 647;
M. Boglione and  M. R. Pennington, Phys. Rev. Lett {\bf 79} (1997) 1998;
N. A. Tornqvist, hep-ph/0008136;
N. A. Tornqvist and A. D. Polosa, Nucl. Phys. {\bf A692} (2001) 259;
A. Deandrea, R. Gatto, G. Nardulli, A. D. Polosa and N. A.
Tornqvist, Phys. Lett. {\bf B502} (2001) 79;
F. De Fazio and M. R. Pennington, Phys. Lett. {\bf B521} (2001) 15;
M. Boglione and  M. R. Pennington, Phys. Rev. {\bf D65}
(2002) 114010 .

\bibitem{CloseKirk00} F. E. Close and A. Kirk, Phys. Lett. {\bf B489} (2000) 24;
A. Kirk, Phys. Lett. {\bf B489} (2000) 29.

\bibitem{CloseKirk01} F. E. Close and A. Kirk, Phys. Lett. {\bf B515} (2001) 13 .

\bibitem{Achasov02} N. N. Achasov and A. Kiselev, Phys. Lett. {\bf B534} (2002) 83 .

\bibitem{VMD02}  D. Black, M. Harada and J. Schechter, Phys. Rev. Lett. {\bf 88} (2002) 181603.


\bibitem{KRS97} O. Krehl, R. Rapp and J. Speth, Phys. Lett. {\bf B390} (1997) 23.

\bibitem{NNAchasov04} N. N. Achasov and  G. N. Shestakov, Phys. Rev. Lett. {\bf 92} (2004) 182001.
\bibitem{Anisovich} V. V. Anisovich, hep-ph/0208123;
V. V. Anisovich, V. A. Nikonov and A. V. Sarantsev, Phys. Atom. Nucl. {\bf 65} (2002) 1545; Yad. Fiz. {\bf65} (2002) 1583;
V. V. Anisovich and A. V. Sarantsev, Eur. Phys. J. {\bf A16} (2003) 229;
 V. V. Anisovich, V. A. Nikonov and A. V. Sarantsev, Phys. Atom. Nucl. {\bf 66} (2003) 741; Yad. Fiz. {\bf66} (2003) 772.


\bibitem{Balitsky} I. I. Balitsky, V. M. Braun and A. V. Kolesnichenko,
Sov. J. Nucl. Phys. {\bf 44}
(1986) 1028 ; Nucl. Phys.  {\bf B312} (1989) 509.

\bibitem{Braun1} V. M. Braun and I. E. Filyanov,
Z. Phys.  {\bf C44} (1989) 157 .

\bibitem{Chernyak1} V. L. Chernyak and I. R. Zhitnitsky, Nucl. Phys. {\bf B345} (1990) 137 .

\bibitem{Shifman79}
 M. A. Shifman, A. I. Vainshtein  and V. I. Zakharov,
 Nucl. Phys. {\bf B147 } (1979) 385, 448.



\bibitem{Belyaev:1994zk}
V. M. Belyaev, V. M. Braun, A. Khodjamirian and R. R\"uckl,
Phys.  Rev. {\bf D51} (1995) 6177.
\bibitem{BK}
J. Bijnens and A. Khodjamirian, Eur. Phys. J. {\bf C26} (2002) 67.

\bibitem{ColangeloFazio2003} P. Colangelo and F. D. Fazio, Phys. Lett. {\bf B559} (2003) 49 .

\bibitem{Yilmaz}  A. Gokalp and O. Yilmaz, Phys. Rev. {\bf D69} (2004) 074023.

\bibitem{Kho1} R. R\"{u}ckl, hep-ph/9810338;
A. Khodjamirian and R. R\"{u}ckl, hep-ph/9801443; V. M. Braun,
hep-ph/9810338.

\bibitem{CZ}
V. L. Chernyak and A. R. Zhitnitsky, Phys. Rep. {\bf 112} (1984) 173.

\bibitem{BF90}
V. M. Braun and I. E. Filyanov, Z. Phys. {\bf C48} (1990) 239.



\bibitem{Ball:1998je}
P. Ball, JHEP {\bf 9901} (1999) 010.

\bibitem{Kim} H. Kim, S. H. Lee and M. Oka, Prog. Theor. Phys. {\bf 109} (2003)
371.


\bibitem{ChengHY} H. Y. Cheng, Phys. Rev. {\bf D67} (2003) 034024.

\bibitem{Anisovich03} V. V. Anisovich, L. G. Dakhno and V. A. Nikonov, hep-ph/0302137.

\bibitem{Aloisio:2002bt}
A. Aloisio {\it et al.}  [KLOE Collaboration],
Phys. Lett.  {\bf B537} (2002) 21.

\bibitem{Akhmetshin:1999di}
R. R. Akhmetshin {\it et al.}  [CMD-2 Collaboration],
Phys. Lett.  {\bf B462} (1999) 380;
Nucl. Phys. {\bf A675} (2000) 424C.

\bibitem{Achasov:2000ym}
M. N. Achasov {\it et al.},
Phys. Lett.  {\bf B485} (2000) 349.

\bibitem{Barberis:1999cq}
D. Barberis {\it et al.}  [WA102 Collaboration],
Phys. Lett.  {\bf B462} (1999) 462.

\bibitem{Gobel:2000es}
E. M. Aitala {\it et al.}  [E791 Collaboration],
Phys. Rev. Lett.  {\bf 86} (2001) 765.

\bibitem{KLOEa0} A. Aloisio et al., KLOE Collaboration, Phys. Lett. {\bf B536} (2002) 209 .

\bibitem{Achasova0} N. N. Achasov and A. V. Kiselev, Phys. Rev. {\bf D68} (2003) 014006  .

\bibitem{Napsuciale:1998ip}
M. Napsuciale,
Hep-ph/9803396;
J. L. Lucio Martinez and M. Napsuciale,
Phys. Lett.  {\bf B454} (1999) 365.

\bibitem{Oller2003} J. A. Oller and E. Oset, Nucl. Phys. {\bf A620} (1997) 438, Erratum-ibid. {\bf A652} (1999)
 407; J. A. Oller, E. Oset and J. R. Pelaez, Phys. Rev. Lett. {\bf 80} (1998)
 3452; Phys. Rev. {\bf D59} (1999) 074001; Erratum-ibid. {\bf D60} (1999) 099906;
 J. A. Oller, Nucl. Phys. {\bf A714} (2003) 161
 ; Nucl. Phys. {\bf A727} (2003) 353.


\bibitem{2004} N. N. Achasov and G. N. Shestakov, hep-ph/0405129; M. Uehara, hep-ph/0404221;
A. V. Anisovich, V. V. Anisovich, V. N. Markov, V. A. Nikonov and  A. V. Sarantsev, hep-ph/0403123.

\end{thebibliography}
\end{document}